\batchmode
\documentstyle[12pt]{article}
\newcommand{\nc}{\newcommand}
\nc{\renc}{\renewcommand}

\nc{\half}{{\textstyle{1\over2}}}
\nc{\etal}{\mbox{\it et al. }}
\nc{\ie}{{\it i.e.}}
\nc{\eg}{{\it e.g.}}

\renc{\thefootnote}{\arabic{footnote}}
\nc{\capt}[1]{{\bf Figure.} {\small\sl #1}}

\nc{\eqs}[2]{\mbox{Eqs.~(\ref{#1},\,\ref{#2})}}
\nc{\eq}[1]{\mbox{Eq.~(\ref{#1})}}

\nc{\figs}[2]{\mbox{Figs.~(\ref{#1},\,\ref{#2})}}
\nc{\fig}[1]{\mbox{Fig~.(\ref{#1})}}

\nc{\tag}[1]{\label{#1} \marginpar{{\footnotesize #1}}}
\nc{\mtag}[1]{\label{#1} \mbox{\marginpar{{\footnotesize #1}}}}
\renc{\baselinestretch}{1.2}
\newlength{\overeqskip}
\newlength{\undereqskip}
\nc{\be}[1]{\begin{equation} \mbox{$\label{#1}$}}
\nc{\bea}[1]{\begin{eqnarray} \mbox{$\label{#1}$}}
\nc{\Section}[2]{\section{#2}\label{#1}}
\nc{\Bibitem}[1]{\bibitem{#1}}
\nc{\Label}[1]{\label{#1}}
\nc{\eea}{\vspace{\undereqskip}\end{eqnarray}}
\nc{\ee}{\vspace{\undereqskip}\end{equation}}
\nc{\bdm}{\begin{displaymath}}
\nc{\edm}{\end{displaymath}}
\nc{\dpsty}{\displaystyle}
\nc{\bc}{\begin{center}}
\nc{\ec}{\end{center}}
\nc{\ba}{\begin{array}}
\nc{\ea}{\end{array}}
\nc{\bab}{\begin{abstract}}
\nc{\eab}{\end{abstract}}
\nc{\btab}{\begin{tabular}}
\nc{\etab}{\end{tabular}}
\nc{\bit}{\begin{itemize}}
\nc{\eit}{\end{itemize}}
\nc{\ben}{\begin{enumerate}}
\nc{\een}{\end{enumerate}}
\nc{\bfig}{\begin{figure}}
\nc{\efig}{\end{figure}}
\nc{\arreq}{&\!=\!&}
\nc{\arrmi}{&\!-\!&}
\nc{\arrpl}{&\!+\!&}
\nc{\arrap}{&\!\!\!\approx\!\!\!&}
\nc{\non}{\nonumber\\*}
\nc{\align}{\!\!\!\!\!\!\!\!&&}

\def\lsim{\; \raise0.3ex\hbox{$<$\kern-0.75em
      \raise-1.1ex\hbox{$\sim$}}\; }
\def\gsim{\; \raise0.3ex\hbox{$>$\kern-0.75em
      \raise-1.1ex\hbox{$\sim$}}\; }
\nc{\DOT}{\hspace{-0.08in}{\bf .}\hspace{0.1in}}
\nc{\Laada}{\hbox {$\sqcap$ \kern -1em $\sqcup$}}
\nc\loota{{\scriptstyle\sqcap\kern-0.55em\hbox{$\scriptstyle\sqcup$}}}
\nc\Loota{{\sqcap\kern-0.65em\hbox{$\sqcup$}}}
\nc\laada{\Loota}
\nc{\qed}{\hskip 3em \hbox{\BOX} \vskip 2ex}

\nc{\real}{{\rm I \! R}}
\nc{\Z}{{\sf Z \!\!\! Z}}
\nc{\complex}{{\rm C\!\!\! {\sf I}\,\,}}
\def\bigid{\leavevmode\hbox{\small1\kern-3.8pt\normalsize1}}
\def\id{\leavevmode\hbox{\small1\kern-3.3pt\normalsize1}}
\nc{\slask}{\!\!\!/}
\nc{\bis}{{\prime\prime}}
\nc{\pa}{\partial}
\nc{\na}{\nabla}
\nc{\ra}{\rangle}
\nc{\la}{\langle}
\nc{\goto}{\rightarrow}
\nc{\swap}{\leftrightarrow}

\nc{\EE}[1]{ \mbox{$\cdot10^{#1}$} }
\nc{\abs}[1]{\left|#1\right|}
\nc{\at}[2]{\left.#1\right|_{#2}}
\nc{\norm}[1]{\|#1\|}
\nc{\abscut}[2]{\Abs{#1}_{\scriptscriptstyle#2}}
\nc{\vek}[1]{{\rm\bf #1}}
\nc{\integral}[2]{\int\limits_{#1}^{#2}}
\nc{\inv}[1]{\frac{1}{#1}}
\nc{\dd}[2]{{{\partial #1}\over{\partial #2}}}
\nc{\ddd}[2]{{{{\partial}^2 #1}\over{\partial {#2}^2}}}
\nc{\dddd}[3]{{{{\partial}^2 #1}\over
	{\partial #2 \partial #3}}}
\nc{\dder}[2]{{{d #1}\over{d #2}}}
\nc{\ddder}[2]{{{d^2 #1}\over{d {#2}^2}}}
\nc{\dddder}[3]{{d^2 #1}\over
	{d #2 d #3}}
\nc{\dx}[1]{d\,^{#1}x}
\nc{\dy}[1]{d\,^{#1}y}
\nc{\dz}[1]{d\,^{#1}z}
\nc{\dl}[1]{\frac{d\,^{#1}l}{(2\pi)^{#1}}}
\nc{\dk}[1]{\frac{d\,^{#1}k}{(2\pi)^{#1}}}
\nc{\dq}[1]{\frac{d\,^{#1}q}{(2\pi)^{#1}}}

\nc{\cc}{\mbox{$c.c.$ }}
\nc{\hc}{\mbox{$h.c.$ }}
\nc{\cf}{cf.\ }
\nc{\erfc}{{\rm erfc}}
\nc{\Tr}{{\rm Tr\,}}
\nc{\tr}{{\rm tr\,}}
\nc{\pol}{{\rm pol}}
\nc{\sign}{{\rm sign}}
\nc{\bfT}{{\bf T }}

\def\GeV{{\rm\ GeV}}

\nc{\cA}{{\cal A}}
\nc{\cB}{{\cal B}}
\nc{\cD}{{\cal D}}
\nc{\cE}{{\cal E}}
\nc{\cG}{{\cal G}}
\nc{\cH}{{\cal H}}
\nc{\cL}{{\cal L}}
\nc{\cO}{{\cal O}}
\nc{\cT}{{\cal T}}
\nc{\cN}{{\cal N}}
\nc{\rvac}[1]{|{\cal O}#1\rangle}
\nc{\lvac}[1]{\langle{\cal O}#1|}
\nc{\rvacb}[1]{|{\cal O}_\beta #1\rangle}
\nc{\lvacb}[1]{\langle{\cal O}_\beta #1 |}
\nc{\bb}{\bar{\beta}}
\nc{\bt}{\tilde{\beta}}
\nc{\ctH}{\tilde{\cal H}}
\nc{\chH}{\hat{\cal H}}
\nc{\1}{\aa}
\nc{\2}{\"{a}}
\nc{\3}{\"{o}}
\nc{\4}{\AA}
\nc{\5}{\"{A}}
\nc{\6}{\"{O}}
\nc{\al}{\alpha}
\nc{\g}{\gamma}
\nc{\Del}{\Delta}
\nc{\e}{\epsilon}
\nc{\eps}{\epsilon}
\nc{\lam}{\lambda}
\nc{\om}{\omega}
\nc{\Om}{\Omega}
\nc{\ve}{\varepsilon}
\nc{\mn}{{\mu\nu}}
\nc{\k}{\kappa}
\nc{\vp}{\varphi}

\nc{\advp}[3]{{\it  Adv.\ in\ Phys.\ }{{\bf #1} {(#2)} {#3}}}
\nc{\annp}[3]{{\it  Ann.\ Phys.\ (N.Y.)\ }{{\bf #1} {(#2)} {#3}}}
\nc{\apl}[3]{{\it  Appl. Phys. Lett. }{{\bf #1} {(#2)} {#3}}}
\nc{\apj}[3]{{\it  Ap.\ J.\ }{{\bf #1} {(#2)} {#3}}}
\nc{\apjl}[3]{{\it  Ap.\ J.\ Lett.\ }{{\bf #1} {(#2)} {#3}}}
\nc{\app}[3]{{\it Astropart.\ Phys.\ }{{\bf #1} {(#2)} {#3}}}
\nc{\cmp}[3]{{\it  Comm.\ Math.\ Phys.\ }{{ \bf #1} {(#2)} {#3}}}
\nc{\cqg}[3]{{\it  Class.\ Quant.\ Grav.\ }{{\bf #1} {(#2)} {#3}}}
\nc{\epl}[3]{{\it  Europhys.\ Lett.\ }{{\bf #1} {(#2)} {#3}}}
\nc{\ijmp}[3]{{\it Int.\ J.\ Mod.\ Phys.\ }{{\bf #1} {(#2)} {#3}}}
\nc{\ijtp}[3]{{\it Int.\ J.\ Theor.\ Phys.\ }{{\bf #1} {(#2)} {#3}}}
\nc{\jmp}[3]{{\it  J.\ Math.\ Phys.\ }{{ \bf #1} {(#2)} {#3}}}
\nc{\jpa}[3]{{\it  J.\ Phys.\ A\ }{{\bf #1} {(#2)} {#3}}}
\nc{\jpc}[3]{{\it  J.\ Phys.\ C\ }{{\bf #1} {(#2)} {#3}}}
\nc{\jap}[3]{{\it J.\ Appl.\ Phys.\ }{{\bf #1} {(#2)} {#3}}}
\nc{\jpsj}[3]{{\it J.\ Phys.\ Soc.\ Japan\ }{{\bf #1} {(#2)} {#3}}}
\nc{\lmp}[3]{{\it Lett.\ Math.\ Phys.\ }{{\bf #1} {(#2)} {#3}}}
\nc{\mpl}[3]{{\it  Mod.\ Phys.\ Lett.\ }{{\bf #1} {(#2)} {#3}}}
\nc{\ncim}[3]{{\it  Nuov.\ Cim.\ }{{\bf #1} {(#2)} {#3}}}
\nc{\np}[3]{{\it  Nucl.\ Phys.\ }{{\bf #1} {(#2)} {#3}}}
\nc{\pr}[3]{{\it Phys.\ Rev.\ }{{\bf #1} {(#2)} {#3}}}
\nc{\pra}[3]{{\it  Phys.\ Rev.\ A\ }{{\bf #1} {(#2)} {#3}}}
\nc{\prb}[3]{{\it  Phys.\ Rev.\ B\ }{{{\bf #1} {(#2)} {#3}}}}
\nc{\prc}[3]{{\it  Phys.\ Rev.\ C\ }{{\bf #1} {(#2)} {#3}}}
\nc{\prd}[3]{{\it  Phys.\ Rev.\ D\ }{{\bf #1} {(#2)} {#3}}}
\nc{\prl}[3]{{\it Phys.\ Rev.\ Lett.\ }{{\bf #1} {(#2)} {#3}}}
\nc{\pl}[3]{{\it  Phys.\ Lett.\ }{{\bf #1} {(#2)} {#3}}}
\nc{\prep}[3]{{\it Phys.\ Rep.\ }{{\bf #1} {(#2)} {#3}}}
\nc{\prsl}[3]{{\it Proc.\ R.\ Soc.\ London\ }{{\bf #1} {(#2)} {#3}}}
\nc{\ptp}[3]{{\it  Prog.\ Theor.\ Phys.\ }{{\bf #1} {(#2)} {#3}}}
\nc{\ptps}[3]{{\it  Prog\ Theor.\ Phys.\ suppl.\ }{{\bf #1} {(#2)} {#3}}}
\nc{\physa}[3]{{\it  Physica\ A\ }{{\bf #1} {(#2)} {#3}}}
\nc{\physb}[3]{{\it  Physica\ B\ }{{\bf #1} {(#2)} {#3}}}
\nc{\phys}[3]{{\it Physica\ }{{\bf #1} {(#2)} {#3}}}
\nc{\rmp}[3]{{\it  Rev.\ Mod.\ Phys.\ }{{\bf #1} {(#2)} {#3}}}
\nc{\rpp}[3]{{\it Rep.\ Prog.\ Phys.\ }{{\bf #1} {(#2)} {#3}}}
\nc{\sjnp}[3]{{\it Sov.\ J.\ Nucl.\ Phys.\ }{{\bf #1} {(#2)} {#3}}}
\nc{\spjetp}[3]{{\it Sov.\ Phys.\ JETP\ }{{\bf #1} {(#2)} {#3}}}
\nc{\yf}[3]{{\it Yad.\ Fiz.\ }{{\bf #1} {(#2)} {#3}}}
\nc{\zetp}[3]{{\it Zh.\ Eksp.\ Teor.\ Fiz.\  }{{\bf #1}  {(#2)} {#3}}}
\nc{\zp}[3]{{\it Z.\ Phys.\ }{{\bf #1} {(#2)} {#3}}}
\nc{\ibid}[3]{{\sl ibid.\ }{{\bf #1} {#2} {#3}}}
\nc{\rf}[1]{(\ref{#1})}
\nc{\nn}{\nonumber \\*}
\nc{\bfB}{\bf{B}}
\nc{\bfv}{\bf{v}}
\nc{\bfx}{\bf{x}}
\nc{\bfy}{\bf{y}}
\nc{\vx}{\vec{x}}
\nc{\vy}{\vec{y}}
\nc{\oB}{\overline{B}}
\nc{\oI}{\overline{I}}
\nc{\oR}{\overline{R}}
\nc{\rar}{\rightarrow}
\nc{\ti}{\times}
\nc{\slsh}{\hskip-5pt/}
\nc{\sm}{Standard~Model~}
\nc{\MP}{M_{\rm Pl}}
\nc{\tp}{t_{\rm Pl}}
\nc{\ave}{\bar{E}}

\nc{\eff}{{\rm eff}}
\nc{\kk}{\vek{k}}
\nc{\pp}{{\rm p}}
\nc{\ga}{g_{a\gamma}}
\nc{\vv}{\\}
\nc{\eee}{{\bf E}}
\nc{\bbb}{{\bf B}}
\nc{\qcd}{T_{\rm QCD}}
\nc{\G}{\rm \ G}
\def\vec#1{{\bf #1}}
\def\lae{\;^{<}_{\sim} \;}  

\def\udd{u^{c}d^{c}d^{c}}

\begin{document}
{\title{\vskip-2truecm{\hfill {{\small HIP-1998-XX/th\\
	 \hfill GUTPA/98/XX/X \\
	}}\vskip 1truecm}
{\bf Observable isocurvature
fluctuations from the Affleck-Dine condensate}}
%\vspace{1.2cm}
{\author{
{\sc  Kari Enqvist$^{1}$}\\
{\sl\small Department of Physics and Helsinki Institute of Physics,}\\ 
{\sl\small P.O. Box 9,
FIN-00014 University of Helsinki,
Finland}\\
{\sc and}\\
{\sc  John McDonald$^{2}$}\\
{\sl\small Department of Physics and Astronomy, University of Glasgow,
Glasgow G12 8QQ,
Scotland}
}
\maketitle
\vspace{1cm}
%\newpage
\begin{abstract}
\noindent

	      In D-term inflation models, Affleck-Dine baryogenesis produces isocurvature 
density fluctuations. These can be perturbations in the baryon number, or, in the case where
 the present neutralino density comes directly from B-ball decay, perturbations in the number
 of 
dark matter neutralinos. The latter case results in a large enhancement of the isocurvature 
perturbation. The requirement that the deviation of the adiabatic perturbations from scale
 invariance due to the Affleck-Dine field is not too large then 
imposes a lower 
bound on the magnitude of the isocurvature fluctuation of about $10^{-2}$ times the 
adiabatic perturbation. This should be observable by MAP and PLANCK. 

\end{abstract}
\vfil
\footnoterule
{\small $^1$enqvist@pcu.helsinki.fi};
{\small $^2$mcdonald@physics.gla.ac.uk}

\thispagestyle{empty}
\newpage
\setcounter{page}{1}
%%%%%%%%%%%%%%%%%%%%%%%%%%%%%%%%%%%%%%%%%%
%%%%%%%%%%%%%%%%%%%%%%%%%%%
The quantum fluctuations of the inflaton field give rise to
fluctuations of the energy density which are adiabatic \cite{eu}.
However, in the minimal supersymmetric standard model (MSSM),
or its extensions, the inflaton is not the only fluctuating field.
It is well known that the MSSM scalar field potential has
many flat directions \cite{drt}, along which a non-zero expectation value can form 
during inflation, leading to a condensate after
inflation, the so-called Affleck-Dine (AD) condensate \cite{ad}.
The AD field is a complex field and, in the currently favoured D-term inflation
models \cite{dti} on which we focus in this letter, 
is effectively massless during inflation. Therefore both its
modulus and phase are subject to fluctuations. In D-term inflation models the phase of the
 AD field receives no 
order $H$ corrections after inflation and so its fluctuations are unsuppressed \cite{kmr}. 
Because the subsequent evolution of the phase of the AD condensate generates the baryon 
asymmetry \cite{ad}, the fluctuations of the phase correspond to 
fluctuations in the local baryon number density, or isocurvature
fluctuations, while the fluctuations of the modulus give rise to adiabatic
density fluctuations. We will show that the adiabatic fluctuations may in fact
 dominate over the
inflaton fluctuations, with potentially adverse consequences for the scale invariance of the 
perturbation spectrum, thus imposing an upper 
bound on the amplitude of the AD field. As a consequence, there is
a lower bound on the isocurvature fluctuation amplitude. The magnitude of this lower bound
 will 
depend on the nature of the AD field. D-term inflation models require that $d > 4$, where
 $d$ 
is the dimension of the non-renormalizable superpotential terms stabilizing the potential, in
 order to avoid thermalizing the AD field too early \cite{kmr}, whilst R-parity conservation,
 required to eliminate dangerous 
renormalizable B and L violating terms from the MSSM, rules out odd values of $d$.
 Therefore we will focus on the $d=6$ direction, in particular the $\udd$ direction, in the
 following. We will show
that the isocurvature fluctuations will most likely be observable in forthcoming 
cosmic microwave background (CMB) satellite experiments.

	An important point is that
the AD condensate is not stable but typically breaks up into non-topological
solitons \cite{ks2,bbb1}
which carry baryon (and/or lepton) number \cite{cole2,ks1}
and are therefore called B-balls (L-balls). This is a generic
feature which is not realized only if the fluctuations take
the AD field into certain leptonic $d=4$ ("$H_{u}L$") directions.
The formation of the B-balls takes place with an efficiency $f_B$, likely
to be in the range 0.1 to 1 \cite{bbb2}.
Hence the AD isocurvature fluctuations are inhereted by the B-balls. 
The properties of the
B-balls depend on SUSY breaking and on the flat direction along which the
AD condensate forms. We will consider SUSY breaking mediated to the observable sector
by gravity. In this case the B-balls are unstable but long-lived,
decaying well after the electroweak phase transition has taken 
place \cite{bbb1}, with a natural order of 
magnitude for decay temperature $T_d\sim {\cal O}(1)\GeV$. 
This assumes a reheating temperature after inflation, $T_R$, of the order of $1 \GeV$. 
Such a low value of $T_R$ is necessary in D-term inflation models because 
the natural magnitude of the phase of the AD field, $\delta_{\rm CP}$, is of the order of 1 
in D-term inflation and along the d=6 direction
 AD baryogenesis implies that the baryon to entropy ratio is $\eta_B\sim \delta_{\rm CP}
 (T_R/10^9\GeV)$ \cite{bbbd}. 
It is significant that a low reheating temperature
 can naturally be achieved in D-term inflation models, as these have discrete symmetries in
 order to ensure 
the flatness of the inflaton potential which can simultaneuously lead to
a suppression of the reheating temperature \cite{bbbd}.

       Because the B-ball is essentially a squark condensate, in R-parity conserving models its
 decay produces 
both baryons and neutralinos ($\chi$), which we assume to be the lightest
supersymmetric particles (LSPs),
with $n_{\chi}\simeq 3n_B$ \cite{bbb2,bbbdm}. This case is particularly interesting, as the
 simultaneous production of baryons and neutralinos may help to explain the 
remarkable similarity of the baryon and dark matter neutralino number densities
 \cite{bbb2,bbbdm}. 
With B-ball decay temperatures $T_d \sim {\cal O}(1)\GeV$,
the decay products no longer thermalize completely and, so long as $T_d$ is
low enough that they do not annihilate after B-ball decay \cite{bbbdm}, retain 
the form of the original AD isocurvature fluctuation. 
Therefore in this scenario the cold dark matter particles
can have both isocurvature and adiabatic density fluctuations, resulting in an enhancement of 
the isocurvature contribution relative to the baryonic case. 
On the other hand, if the neutralinos from B-ball decays annihilate, 
the neutralino contribution to the isocurvature fluctuation will be erased, leaving 
only the baryonic contribution. Although we will be primarily interested in the neutralino 
isocurvature fluctuation case, we will also comment on the purely baryonic case in the
 following.

	Isocurvature perturbations have been studied previously \cite{stomper},
in particular in the context of axion models \cite{axion,burns}.
The isocurvature perturbations give rise
to extra power at large angular scales but are damped at small angular
scales. The 
amplitude of the rms mass fluctuations in an $8h^{-1}$ Mpc$^{-1}$
sphere, denoted as $\sigma_8$, is about an order of magnitude lower 
than in the adiabatic case. Hence COBE normalization alone is
sufficient to set a tight limit on the relative strength of
the isocurvature amplitude. Small isocurvature fluctuations
are, however, beneficial, in that they improve the
fit to the power spectrum in $\Omega_{0} = 1$ 
CDM models with a cosmological constant \cite{axion}
(or $\Omega_{0} = 1$, $\Lambda = 0$  
CDM models with some hot dark matter \cite{burns}). 
For instance, in the context
of axion models it has been found \cite{axion} that an $\Omega_{0} = 1$ 
mixed fluctuation model with a relative isocurvature perturbation
amplitude of 5\%, $\Omega_{a}=0.4$ and $\Omega_{\Lambda} = 0.6$
would give a very good fit to the data. However, the 
 isocurvature fluctuations seem
to require a large axion decay constant, which is already excluded
unless there is considerable late entropy production \cite{axion}.
The Affleck-Dine case we consider here is more economic, in the sense that it requires only
the particles of the MSSM. 

	     In D-term inflation models, the AD field 
$\Phi=\phi e^{i\theta}/\sqrt{2}\equiv (\phi_1+i\phi_2)/\sqrt{2}$ 
remains effectively massless during inflation. Therefore the real fields $\phi_i$
are subject to quantum fluctuations with
\be{fluct}
\delta \phi_i(\vec x)=\sqrt{V}\int {d^3k\over (2\pi)^3}e^{-i\vec k\cdot\vec x}
\delta_{\vec k}~~,
\ee
where $V$ is a normalizing volume and where the power spectrum is the same as for the
 inflaton field,
\be{scaleinv}
{k^3\vert\delta_{\vec k}\vert^2\over 2\pi^2}=\left({H_I\over 2\pi}\right)^2~~,
\ee
where $H_I$ is the value of the Hubble parameter during inflation. Thus,
for given background values $\bar\theta$ and $\bar\phi$, 
(with $\bar\theta$ naturally of the order of 
1) one finds 
\be{thetapert} \left({\delta
 \theta\over\;Tan(\bar\theta)}\right)_k={H_I\over\;Tan(\bar\theta)\bar\phi}
={H_Ik^{-3/2}\over\sqrt{2}Tan(\bar\theta)\bar\phi_I}~~,
\ee
where $\phi_I$ is the value of $\phi$ when the perturbation leaves the
horizon. After inflation, during the inflaton oscillation dominated period, the mass 
squared of the magnitude of the AD field will receive an order $H^2$ correction, which 
must be negative in order to have a non-zero $\phi$ and so AD baryogenesis \cite{drt},
 whilst its phase receives no order $H$ corrections. 
Therefore, the magnitude of the AD field $\Phi$ remains at the non-zero minimum of its
 potential until 
$H\simeq m_S$, where $m_S \sim 100 \GeV$ is the SUSY breaking scalar mass,
whence it begins to oscillate and the baryon asymmetry
$n_B \propto Sin(\theta)$ forms. Since $\bar\theta$ and $\delta\theta$ remain
constant until $H\simeq m_S$, we have
\be{nb}
\left({\delta n_B\over n_B}\right)_k
=\left({\delta \theta\over\; Tan(\bar\theta)}\right)_k
~~
\ee
with ${\delta \theta/Tan(\bar\theta)}$ given by \eq{thetapert}.

	We first consider the case where the adiabatic perturbation is mostly due to the inflaton. 
The adiabatic perturbation is determined by the invariant
$\zeta = \delta\rho/(\rho+p)$ with $\delta\rho=V'\delta\phi$.
During inflation, when all the fields are slow rolling, one finds \cite{burns}
\be{adi}
\zeta_{adiab}=\frac 34\delta_\gamma^{(a)}
={9\over \sqrt{2}}{H_I^3\over V'}k^{-3/2}~~,
\ee
where $ \delta_{\gamma} \equiv \delta \rho_{\gamma}/\rho_{\gamma}$. 
For super-horizon size isocurvature fluctuations $\delta \rho/\rho=0$,
 so that $m_\chi\delta n_\chi+m_B\delta n_B+4(\rho_\gamma+\rho_\nu)\delta T/T
=0$ (here $\rho_\gamma$ and $\rho_\nu\simeq 0.68\rho_\gamma$ are
respectively the photon and the neutrino densities, and we assume 
for simplicity that there are no massive neutrinos).
We then find that in the presence of both
adiabatic ($\delta(n_x/s) =0$) 
and isocurvature ($\delta(n_x/s) \ne 0$) fluctuations
\be{delt}
{\delta T\over T}=-{\rho_\chi\delta^{(i)}_\chi+\rho_B\delta^{(i)}_B\over
3(\rho_\chi+\rho_B)+4(\rho_\gamma+\rho_\nu)}~~,
\ee
where $\delta_{x} = \delta n_{x}/n_{x}$ for non-relativistic particles $x$. 
The isocurvature fluctuations of the LSPs are related to the
baryonic isocurvature fluctuations by $\delta n_\chi^{(i)}=3f_B
\delta n_B^{(i)}$, with $\delta n_B^{(i)}$ given by \eq{nb}.
In the linear perturbation theory adiabatic and isocurvature
fluctuations evolve independently so that the total perturbation
is just the sum of the two.

      The total LSP number density is the sum of the thermal relic density
$n_\chi^{(th)}$ and the density $n_\chi^{(B)}=3f_Bn_B$
 originating from the B-ball decay.
Using \eq{delt}, 
the isocurvature fluctuation
imposed on the CMB photons is then found to be
\bea{delg}
\delta_\gamma^{(i)}&=&4{\delta T\over T}=-{4 \left(1+{m_B\over 3f_Bm_\chi}\right)
\rho_\chi^{(B)}\delta^{(i)}_B
\over 3(\rho_\chi+\rho_B)+4(\rho_\gamma+\rho_\nu)}\nn
&\simeq& -\frac 43\left(1+{m_B\over 3f_Bm_\chi}\right)
\left({\Omega_\chi-\Omega_\chi^{(th)}\over \Omega_m}\right)\delta^{(i)}_B
\equiv -\frac 43\omega\delta^{(i)}_B~~,
\eea
where $\rho_\chi^{(B)}$ is the LSP mass density from the B-ball, 
$\Omega_m~(\Omega_\chi)$ is total matter (LSP) 
density (in units of the critical density), and $\delta^{(i)}_B$
is given by \eq{nb}. To obtain the last
line in \eq{delg}, we have used the fact that $\rho_\gamma$ is negligible.
In the notation of reference \cite{burns} and using \eq{adi} we can write 
\be{bdef}
\beta\equiv \left( {\delta_\gamma^{(i)}\over \delta_\gamma^{(a)}}\right)^2
= \frac 19 \omega^2
\left({M^2V'(S)\over V(S)Tan(\bar\theta)\bar\phi}\right)^2~~,
\ee
where $S$ is the inflaton field with a potential $V(S)$ and
$M\equiv M_{Pl}/\sqrt{8\pi}$. 

	  In the simplest D-term inflation model,
the inflaton is coupled to the matter fields $\psi_-$ and $\psi_+$ 
carrying opposite  Fayet-Iliopoulos charges through a superpotential
term $W=\kappa S\psi_-\psi_+$ \cite{dti,kmr}. At one loop level the inflaton potential
reads
\be{infpot}
V(S)=V_0+{g^4\xi^4\over 32\pi^2}\ln\left({\kappa^2S^2\over Q^2}\right)
\;\; ; \;\;\;\; V_0 = \frac{g^2 \xi^4}{2}       ~~,\ee
where $\xi$ is the Fayet-Iliopoulos term and $g$ the gauge coupling 
associated with it. COBE normalization fixes $\xi=6.6\times 10^{15}\GeV$ \cite{lr}. 
In addition, we must consider the contribution of the AD field to the 
adiabatic perturbation. During inflation, the potential of the $d=6$ flat AD field
is simply given by
\be{adpot}
V(\phi)={\lambda^2\over 32 M^6}\phi^{10}~~.
\ee
With $\rho=V(S)+V(\phi)$ and
$\rho+p=\dot S^2+\dot\phi^2$ one finds, taking both $S$ and
$\phi$ to be slow rolling fields with $\dot S=-V'(S)/(3H_I)$
and $\dot\phi=-V'(\phi)/(3H_I)$, that the invariant $\zeta$ is now
\be{adiab}
\zeta_{adiab}\propto {V'(\phi)+V'(S)\over V'(\phi)^2+V'(S)^2} \; \delta \phi ~~,
\ee
where we have used the fact that both fields are massless,
so that $\delta S=\delta \phi$. 
Thus the field which dominates the spectral index of the perturbation will be that 
with the largest value of $V^{'}$ and $V^{''}$. 
The index of the power spectrum 
is given by $n=1+2\eta-6\epsilon$, 
where $\epsilon$ and $\eta$ are defined as
\be{defs}
\epsilon=\frac 12{M^2}\left({V'\over V}\right)^2~~,~~
\eta={M^2}{V''\over V}~~.
\ee
The present
lower bounds imply $|\Delta n| \lsim 0.2$. (This bound will be much improved by future 
satellite missions). 
In the case where the derivatives with respect to the  
inflaton dominate (for which the potential is dominated by 
$V_{0}$ for all $\xi < M$), $|\Delta n| = 1/N \approx 0.02$ for $N \sim 50$. Once the
 derivatives with respect to the AD field dominate, the spectral index increases rapidly with
 $\phi$; from $\eta$ 
($\epsilon$), $|\Delta n|$ is proportional to $\phi^{8}$ ($\phi^{18}$). The condition for the 
AD field to dominate the spectral index is that $\phi > Max(\phi_{c_1}, \phi_{c_2})$, where
\be{phi1}
\phi_{c_1}\simeq 0.64 (g^3\lambda^{-2}\xi^4M^5)^{1/9}~~
\ee
 and 
\be{phi2}
\phi_{c_2}\simeq 0.48 \left(\frac g\lambda\right)^{1/4}(M \xi)^{1/2}~~,
\ee
and where we have used the fact that during the slow roll-over $S_{N}^2\simeq
 g^2M^2N/(4\pi^2)$.
The inflaton derivatives will dominate once $\phi < Min(\phi_{c_1}, \phi_{c_2})$. 
(In practice $\phi_{c_1}$ and $\phi_{c_2}$ only differ by a factor of less than 2). 
As a result of the rapid increase of the spectral index once the AD derivatives dominate, the
 condition that the spectral index is acceptably close to 
scale invariance essentially reduces to the condition that it is dominated 
by the inflaton. The lower bounds on $\beta$ corresponding to $\phi_{c_{1}}$ and
 $\phi_{c_{2}}$ are then 
\be{blimit1}
\beta>\beta_{c_1}\simeq
6.5\times 10^{-3}g^{4/3}\lambda^{4/9}\omega^2 Tan(\bar\theta)^{-2}~~
\ee
and 
\be{blimit2}
\beta>\beta_{c_2}=2.5 \times 10^{-2}g^{3/2}\lambda^{1/2}\omega^2
Tan(\bar\theta)^{-2}~~.
\ee
(For most values of the couplings the latter leads to a slightly more stringent lower bound). 
Thus significant isocurvature fluctuations are a definite prediction of the AD mechanism.

%%%%%%%%%%%%%%%%%%%%%%%%%%%%%%%%%%%%%%%%%%
%%%%%%%%%%%%%%%%%%%%%%%%%%%%%%%%
\begin{figure}
\leavevmode
\centering
\vspace*{90mm} 
\includegraphics{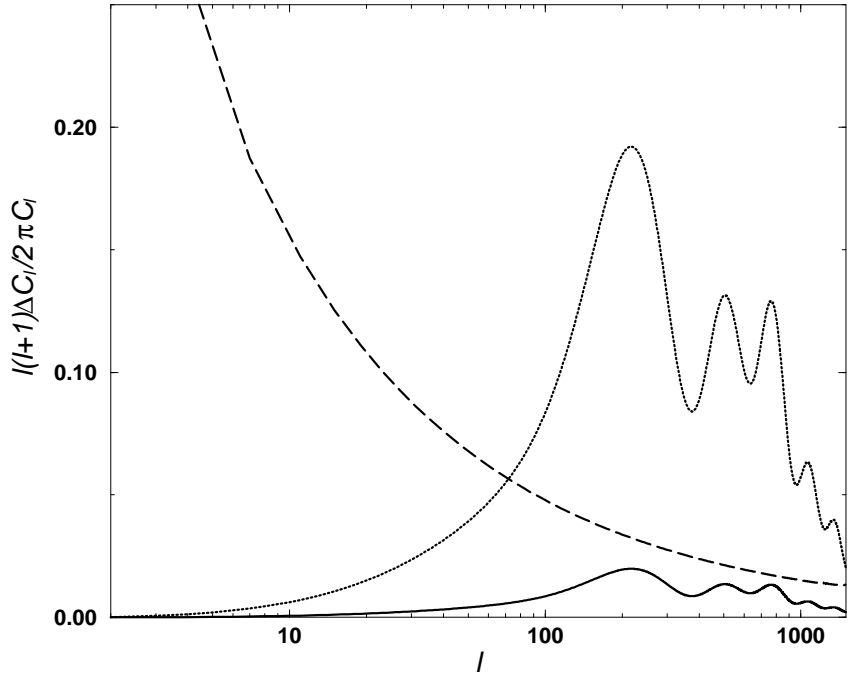}
%\special{psfile=johnqbpot.ps voffset=+30 hoffset=-5 vscale=35 hscale=35
%angle=270}   
\caption{The relative difference $\Delta C_l/C_l$
between the purely adiabatic and 
a mixture of adiabatic and isocurvature angular power spectra with
$\beta=0.001$ (dotted line) and
$\beta=0.0001$ (solid line) for a purely
CDM $\Omega=1$ model (with $\Omega_B=0.05$, $h=0.5$
and the spectral index $n=1$).
Shown is also the projected PLANCK error level, averaged over
ten multipoles (dashed line).}
\label{kuva2}       
\end{figure} 

%%%%%%%%%%%%%%%%%%%%%%%%%%%%%%%%%%%%%%%%%%
%%%%%%%%%%%%%%%%%%%%%%%%%%%%%%%%

	     There are two limiting cases: $f_{B} \gg m_{B}/3f_{B}m_{\chi}$, for which
 $\omega = 1$, and $f_{B} \ll m_{B}/3f_{B}m_{\chi}$, for which $\omega =
 \Omega_{B}/\Omega_{m}$. The latter corresponds to the case where the B-balls form
 very inefficiently or where the neutralino contribution to the isocurvature perturbation is
 erased by annihilations \cite{bbb2,bbbdm}. In this case only the baryonic isocurvature
 perturbation remains. 

	       The actual lower limit on $\beta$ depends on the unknown couplings
$g$ and $\lambda$, as well as on $\bar\theta$. To obtain an estimate
for $\beta_{c_2}$ in the case where the dark matter neutralinos come directly from B-ball
 decay, let us adopt the following values: $g\simeq g_{\rm GUT} \simeq 0.7$,
 $\lambda\simeq 1/5!=0.008$ (corresponding to non-renormalizible interaction 
with physical strength set by $M$ \cite{kmr,bbbd}), $Tan(\bar\theta)^2\simeq 1$
 (corresponding to $\bar \theta \approx \pi/4 $) and, assuming $f_B$ is not too small,  
$\omega \simeq 1$. We then find that $\beta>\beta_{c_2}\simeq 1.3 \times
10^{-3}$, with the conservative lower bound perhaps an order of magnitude
smaller, $ \beta \sim 10^{-4}$. For the purely baryonic case, the value of $\omega$ 
depends on $\Omega_{B}$ and $\Omega_{m}$. Nucleosynthesis combined with 
the current best estimate of the expansion rate ($0.6 \lae h \lae 0.87$ \cite{exp}) implies that 
$0.006 \lae \Omega_{B} \lae 0.036$. Thus with $\Omega_{m} = 1 \; (0.4)$ 
we obtain that $\omega$ for neutralino case is 30-150 (10-60) times larger 
than in the purely baryonic case. Therefore in the baryonic case the corresponding value of
 $\beta$ is two to 
four orders of magnitude smaller. 
Thus there is a significant enhancement of the isocurvature perturbation in the case where 
the dark matter neutralinos come from B-ball decay without subsequent annihilations. 
It should be emphasized that there is no physical reason to expect 
$\phi$ to be close to its upper bound, so $\beta$ may, in general, be expected to be much
 larger that these lower bounds. In this case, even without the neutralino enhancement, the
 purely baryonic 
isocuvature fluctuation may well be important.

	      In Fig. 1 we display the difference between the purely adiabatic 
power spectrum and the spectra with $\beta \ne 0$. We also plot
the expected error for the Planck Surveyor Mission, following
the estimates in ref. \cite{bond}. (A similar error is expected for MAP for 
$l \lsim 500$). 
The standard error reads
$(\Delta C_l)^2=2(C_l+\delta)^2/[(2l+1)f_{\rm sky}]$, where
$f_{\rm sky}$ is the fraction of the sky sampled
(we take $f_{\rm sky}=0.65$) and $\delta$ is from the beam, the
angular resolution and the sensitivity, as discussed in  \cite{bond}; 
$\delta$ becomes non-negligible only for $l\gsim 1000$ for PLANCK and 
$l \gsim 500$ for MAP.
One should bear in mind that, in principle, each multipole provides
an independent measurement of the spectrum.
As can be seen, detecting isocurvature fluctuations at the level
of $\beta\sim 10^{-4}$ should be quite realistic by averaging over
a sufficient number of multipole measurements. However, detecting
or setting an actual lower limit on $\beta$ will require a much more careful analysis,
which we do not attempt here. 
Nevertheless, on the basis of
Fig. 1, it seems likely that the forthcoming CMB experiments will 
definitely be able to see
isocurvature perturbations in the case where the baryons and neutralinos 
come directly from the decay 
of unstable B-balls in the context of D-term inflation models, 
hence offering a test not only
of the inflationary Universe but also of the B-ball variant of AD baryogenesis.

		  In conclusion, AD baryogenesis in the context of D-term inflation generally 
implies the existence of isocurvature density fluctuations. In the case where B-balls, which 
are generally expected to form in AD baryogenesis, decay late enough to produce the
 observed neutralinos without annihilations, the isocurvature fluctuctions should be observable
 by MAP and PLANCK. Even in the case where only baryonic isocurvature fluctuations
 arise, there is still a 
reasonable possibility of observing them, although in this case it is less 
certain. Thus isocurvature fluctuations are a clear fingerprint of D-term inflation and 
Affleck-Dine baryogenesis. In particular, for the case where the neutralino dark matter comes
 directly from B-ball decays, which allows for an understanding of the remarkable similarity of
 the baryon and dark matter number densities \cite{bbb2,bbbdm}, observation of
 isocurvature perturbations combined with the non-thermal nature of the dark matter
 neutralino density (testable by observation of the sparticle spectrum \cite{bbbdm}) would
 strongly support the late decaying B-ball scenario and D-term inflation, giving us a deep
 insight into the nature of particle physics and the very early Universe.

\subsection*{Acknowledgements}   
We would like to acknowledge the use of CMBFAST to calculate
the angular power spectrum.
This work has been supported by the
 Academy of Finland  under the contract 101-35224
and by a European Union Marie Curie Fellowship under EU contract number 
ERBFM-BICT960567.

\end{document}